\documentclass[12pt]{article}
\usepackage{amsfonts}
\usepackage{amsmath}
\usepackage{amssymb}

\input{tcilatex}
\begin{document}

\begin{center}
\smallskip \ 

\textbf{TOWARDS AN ALTERNATIVE GRAVITATIONAL THEORY}

\smallskip \ 

\bigskip \ 

J. A. Nieto \footnote{%
nieto@uas.edu.mx, janieto1@asu.edu}, L. A. Beltr\'{a}n

\smallskip \ 

\smallskip \ 

\smallskip

\textit{Facultad de Ciencias F\'{\i}sico-Matem\'{a}ticas de la Universidad}

\textit{Aut\'{o}noma} \textit{de Sinaloa, 80010, Culiac\'{a}n, Sinaloa, M%
\'{e}xico.}

\smallskip \ 

\smallskip \ 

\textbf{Abstract}
\end{center}

In 1680 Cassini proposed oval curves as alternative trajectories for the
visible planets around the sun. The Cassini ovals were of course overshadow
by the Kepler's first law (1609), namely the planets move around the sun
describing conic orbits. Here we describe the possibility that the Cassini's
idea works at larger or smaller scales. Indeed, we consider the Spiric
curves (which are a generalization of the Cassini oval) and present the
first steps towards a Spiric gravitational theory. We show that from our
formalism an ellipse associated with a planet can be obtained as a
particular case.

\bigskip \ 

\bigskip \ 

\bigskip \ 

\bigskip \ 

\bigskip \ 

\bigskip \ 

\bigskip \ 


Pacs numbers: 04.20.Gz, 04.60.-Ds, 11.30.Ly

May 13, 2014

\newpage

It is a fact that quantum theory of the gravitational field is still an open
problem (see Refs. [1]-[2] and references therein). The two main proposals
superstrings theory [3] and loop quantum gravity theory [4] are valuable
attempts but contains many difficulties. For instance, superstring theory
requires to compactify a number of the higher dimensions in which the theory
is consistent [3]. While loop quantum gravity is, in general, still forced
to `live' in four dimensions, but there are serious problems in specifying
the meaning of the time evolution for the quantum states [5]-[6]. There are
attempts to make sense of loop quantum gravity in higher dimensions [7]-[8],
but the central concept in four dimensions of duality seems to be lost.
(However, it is interesting that, by means of the octonion structure, the
duality concept may still be recasted in eight dimensions [9]-[10].) One may
view such a difficulties as an opportunity to look for other unexpected
alternatives for quantum gravity. In this context, one may review
traditional starting points in the construction of the gravitational field.
The idea is to revisit the Newton universal gravitational theory itself.
This theory is based on the potential

\begin{equation}
V=-\frac{k}{r},  \tag{1}
\end{equation}%
where the constant $k=GMm_{0}$ is given in terms of the Newton gravitational
constant $G$ and two masses $M$ and $m_{0}$ which are separated a distance
determined by the radius $r$. Of course, when (1) is considered in the
Newton second law the conic orbits for the planets are obtained. Here, we
explore the possibility that (1) may be generalized even in the
non-relativistic case through slicing the torus instead of the cone.

Our strategy will be to assume that, instead of the conic orbits, an object
of mass $m_{0}$ moves around another of mass $M$ describing a generalized
version of the Cassini oval curve [11]-[12]. Such generalization refers to
the so called Spiric sections (the word Spiric means torus in ancient
Greek). It turns out that the Spiric sections are an extension of the conic
sections constructed by Menaechmus around 150 BC by cutting a cone by a
plane, and were first considered around 50 AD by the Greek mathematician
Perseus. These Spiric curves are obtained when instead of cutting a cone one
slices a torus with a plane. And in fact, they are assumed to be the first
toric sections to be described. We show that from our formalism the ellipse
curve can be obtained as a particular case. Moreover, we attempt to
generalize the formalism to the relativistic scenario. We argue that the
Spiric sections may eventually lead to a gravitational theory with a
four-rank metric [13]-[18]. We comment about the possible geometric object
implied by such a four-rank metric such as a generalized eight-rank Riemann
curvature tensor. We also explore the possible implications of our formalism
in string theory and other physical scenarios.

Let us start recalling the Lagrangian associated with (1) [19], namely

\begin{equation}
L=\frac{1}{2}m_{0}(\dot{r}^{2}+r^{2}\dot{\theta}^{2})+\frac{k}{r}.  \tag{2}
\end{equation}%
Here, the dot means derivative with respect a time-like parameter. From (2)
one obtains the equations of motion

\begin{equation}
m_{0}\ddot{r}-m_{0}r\dot{\theta}^{2}+\frac{k}{r^{2}}=0,  \tag{3}
\end{equation}%
and%
\begin{equation}
\frac{d}{dt}(m_{0}r^{2}\dot{\theta})=0.  \tag{4}
\end{equation}%
The formula (4) implies

\begin{equation}
m_{0}r^{2}\dot{\theta}=l,  \tag{5}
\end{equation}%
where $l$ is constant of the motion, which eventually may be related with
the orbital angular momentum. The substitution of (5) into (3) leads to

\begin{equation}
m_{0}\ddot{r}-\frac{l^{2}}{m_{0}r^{3}}+\frac{k}{r^{2}}=0.  \tag{6}
\end{equation}%
It is well known that starting from (6) one may derive the orbit expression
[19],%
\begin{equation}
r=\frac{a(1-\varepsilon ^{2})}{1+\varepsilon \cos \theta }.  \tag{7}
\end{equation}%
Here, $a$ and $\varepsilon $ are constant which can be identified with the
semi major axis and the eccentricity of the orbit, respectively.

Now, we are interested in thinking backwards in the sense of assuming first
the orbit (7) and then by using (5) to derive (6). When this computation is
realized one proves that in fact (7) implies (6). With (6) at hand one may
now proceed to write the Lagrangian (2).

Let us now assume that instead of (7) one has the orbit,

\begin{equation}
r^{4}-2d^{2}r^{2}\cos 2\theta +\alpha ^{2}r^{2}+\beta ^{4}=0,  \tag{8}
\end{equation}%
where $\alpha $ and $\beta $ are constants. One notice that this expression
looks very different than (7). It turns out that when $\alpha =0$ (8)
corresponds to the Cassini oval and if in addition $\beta =0$ one obtains
famous Bernoulli lemniscate. Our first goal here, it is to begin with (8)
and then to derive the analogue of (6).

For this purpose let us first write (8) as

\begin{equation}
r^{2}+\frac{\beta ^{4}}{r^{2}}+\alpha ^{2}=2d^{2}\cos 2\theta .  \tag{9}
\end{equation}%
By taking the derivative (with respect the time) of this expression and
using (5) one obtains

\begin{equation}
\dot{r}=-\frac{l}{m_{0}}r^{-1}(r^{2}-\frac{\beta ^{4}}{r^{2}}%
)^{-1}(4d^{4}-(r^{2}+\frac{\beta ^{4}}{r^{2}}+\alpha ^{2})^{2})^{1/2}. 
\tag{10}
\end{equation}%
Deriving (10) once again and using (10) itself yields%
\begin{equation}
\ddot{r}=\frac{l^{2}}{m_{0}r^{3}}-A\frac{(3r^{2}+\frac{\beta ^{4}}{r^{2}})}{%
r^{3}(r^{2}-\frac{\beta ^{4}}{r^{2}})^{3}}+B\frac{(r^{4}+3\beta ^{4})}{%
r^{3}(r^{2}-\frac{\beta ^{4}}{r^{2}})^{3}},  \tag{11}
\end{equation}%
where

\begin{equation}
A=\frac{4l^{2}(d^{4}-\beta ^{4}-\frac{\alpha ^{4}}{4})}{m_{0}^{2}}  \tag{12}
\end{equation}%
and%
\begin{equation}
B=\frac{4l^{2}\alpha ^{2}}{m_{0}^{2}}.  \tag{13}
\end{equation}%
It is not difficult to see that if

\begin{equation}
V_{1}=-\frac{A}{2r^{2}(r^{2}-\frac{\beta ^{4}}{r^{2}})^{2}}  \tag{14}
\end{equation}%
one gets

\begin{equation}
\frac{\partial V_{1}}{\partial r}=\frac{A(3r^{2}+\frac{\beta ^{4}}{r^{2}})}{%
r^{3}(r^{2}-\frac{\beta ^{4}}{r^{2}})^{3}},  \tag{15}
\end{equation}%
and if 
\begin{equation}
\begin{array}{c}
V_{2}=B[-\frac{1}{(r^{2}-\frac{\beta ^{4}}{r^{2}})^{2}}+\frac{1}{2\beta ^{2}}%
\{ \frac{1}{2(r^{2}-\beta ^{2})}+\frac{1}{4(r^{2}-\beta ^{2})^{2}} \\ 
\\ 
-\frac{1}{2(r^{2}+\beta ^{2})}+\frac{\beta ^{2}}{4(r^{2}+\beta ^{2})^{2}}\}],%
\end{array}
\tag{16}
\end{equation}%
one finds

\begin{equation}
\frac{\partial V_{2}}{\partial r}=\frac{B(r^{4}+3\beta ^{4})}{r^{3}(r^{2}-%
\frac{\beta ^{4}}{r^{2}})^{3}}.  \tag{17}
\end{equation}%
Thus, if $\mathcal{V}=V_{1}+V_{2}$, that is, if

\begin{equation}
\begin{array}{c}
\mathcal{V}=-\frac{A}{2r^{2}(r^{2}-\frac{\beta ^{4}}{r^{2}})^{2}}+B[-\frac{1%
}{(r^{2}-\frac{\beta ^{4}}{r^{2}})^{2}}+\frac{1}{2\beta ^{2}}\{ \frac{1}{%
2(r^{2}-\beta ^{2})}+\frac{1}{4(r^{2}-\beta ^{2})^{2}} \\ 
\\ 
-\frac{1}{2(r^{2}+\beta ^{2})}+\frac{\beta ^{2}}{4(r^{2}+\beta ^{2})^{2}}\}],%
\end{array}
\tag{18}
\end{equation}%
then one finds that (11) can be written as%
\begin{equation}
\ddot{r}=\frac{l^{2}}{m_{0}r^{3}}-\frac{\partial \mathcal{V}}{\partial r}. 
\tag{19}
\end{equation}%
It is evident that the potential (18) looks very different than the
Newtonian potential (1). It is intriguing that even in the non relativistic
case when $r=\beta $ the potential (18) determines a singularity, analogue
to the singularity in black-holes.

A natural question is to understand the geometry behind (8). In fact, one
can show that (8) can be obtained from the parametric torus equations

\begin{equation}
\begin{array}{c}
x=(d+a\cos u)\cos \phi , \\ 
\\ 
y=(d+a\cos u)sen\phi , \\ 
\\ 
z=bsenu.%
\end{array}
\tag{20}
\end{equation}%
Here, $a,b$ and $d$ are constants associated with the torus geometry. From
(20) one first finds the formula%
\begin{equation}
b^{2}(\sqrt{x^{2}+y^{2}}-d)^{2}+a^{2}z^{2}-a^{2}b^{2}=0,  \tag{21}
\end{equation}%
which leads to%
\begin{equation}
(b^{2}y^{2}+a^{2}z^{2}+b^{2}x^{2}+b^{2}d^{2}-a^{2}b^{2})^{2}-4b^{4}d^{2}(y^{2}+x^{2})=0.
\tag{22}
\end{equation}%
In turn, this equation can be rewritten in the interesting alternative form:

\begin{equation}
\begin{array}{c}
b^{4}x^{4}+2b^{2}[b^{2}(y^{2}-d^{2})+a^{2}(z^{2}-b^{2})]x^{2} \\ 
\\ 
+[b^{2}(y-d)^{2}+a^{2}(z^{2}-b^{2})][b^{2}(y+d)^{2}+a^{2}(z^{2}-b^{2})]=0.%
\end{array}
\tag{23}
\end{equation}%
Giving different values to $x$ one can obtain various well known curves. For
instance if $x=0$ then (23) leads to two ellipses: at distances $d$ and $-d$%
, respectively. If $a=b=x\neq 0$ then by choosing polar coordinates for $y$
and $z$ one can prove that (23) yields to (8), with $\alpha =d^{2}$ and $%
\beta =4a^{2}d^{2}$. Moreover, if in addition one assumes $d=2a$ one ends up
with the Bernoulli lemniscate. So, it seems that torus constraint (23) is
the right route beyond the Cassini oval. Furthermore, the various values of $%
x$ determine several slices of the torus which in mathematics are called
Spiric sections.

Our next step will be to make the constraint (23) relativistic. For this
purpose one can assume that just as the constraint%
\begin{equation}
r_{0}^{2}=x^{2}+y^{2}+z^{2},  \tag{24}
\end{equation}%
is generalized first to the relativistic form

\begin{equation}
s^{2}=-c^{2}t^{2}+x^{2}+y^{2}+z^{2},  \tag{25}
\end{equation}%
and then infinitesimally as%
\begin{equation}
ds^{2}=-c^{2}dt^{2}+dx^{2}+dy^{2}+dz^{2},  \tag{26}
\end{equation}%
the quartic equation (23), in a proper choosing basis, may be extended to
the form

\begin{equation}
ds^{4}=\xi _{2}(-c^{2}dt^{2}+dx^{2}+dy^{2}+dz^{2})+\xi
_{4}(-c^{4}dt^{4}+dx^{4}+dy^{4}+dz^{4}).  \tag{27}
\end{equation}%
Here, in order to have the right units, we introduced to constant factors $%
\xi _{2}$ and $\xi _{4}$. In a covariant form this expression can be written
as

\begin{equation}
ds^{4}=\xi _{2}\eta _{\mu \nu }dx^{\mu }dx^{\nu }+\xi _{4}\eta _{\mu \nu
\alpha \beta }dx^{\mu }dx^{\nu }dx^{\alpha }dx^{\beta }.  \tag{28}
\end{equation}

Following with the analogy, one may now generalize (28) to a curved
space-time as

\begin{equation}
ds^{4}=\xi _{2}g_{\mu \nu }(x^{\sigma })dx^{\mu }dx^{\nu }+\xi _{4}g_{\mu
\nu \alpha \beta }(x^{\sigma })dx^{\mu }dx^{\nu }dx^{\alpha }dx^{\beta }. 
\tag{29}
\end{equation}%
Furthermore, an electromagnetic potential $g_{\mu }\equiv A_{\mu }$ and a
new kind of generalized potential $g_{\mu \nu \alpha }(x^{\sigma })\equiv
A_{\mu \nu \alpha }(x^{\sigma })$ may be considered in the form

\begin{equation}
\begin{array}{c}
ds^{4}=\xi _{1}g_{\mu }(x^{\sigma })dx^{\mu }+\xi _{2}g_{\mu \nu }(x^{\sigma
})dx^{\mu }dx^{\nu }+\xi _{3}g_{\mu \nu \alpha }(x^{\sigma })dx^{\mu
}dx^{\nu }dx^{\alpha } \\ 
\\ 
+\xi _{4}g_{\mu \nu \alpha \beta }(x^{\sigma })dx^{\mu }dx^{\nu }dx^{\alpha
}dx^{\beta }.%
\end{array}
\tag{30}
\end{equation}%
Of course (in analogy to the electric charge), $\xi _{1}$ and $\xi _{3}$
play the role of constant charges. The expression (30) suggests the
straightforward generalization

\begin{equation}
\begin{array}{c}
ds^{2n}=\xi _{1}g_{\mu _{1}}(x^{\sigma })dx^{\mu _{1}}+\xi _{2}g_{\mu
_{1}\mu _{2}}(x^{\sigma })dx^{\mu _{1}}dx^{\mu _{2}} \\ 
\\ 
+...+\xi _{n}g_{\mu _{1}...\mu _{n}}(x^{\sigma })dx^{\mu _{1}}...dx^{\mu
_{n}}.%
\end{array}
\tag{31}
\end{equation}%
Just as $ds^{4}$ is connected with a torus of genus $n=1$, it seems that $%
ds^{4n}$ may be related to the Riemann surface of genus $n$.

Here, however it seems to us that one need a criteria to set exceptional
cases. Requiring conformal invariant geometrical system in higher energy may
be a possibility [14]. Another possibility is to consider an extension of
the Hurwitz theorem since it is known that division algebras over the real
exist only in dimensions $1,2,4$ and $8$ [20]-[24]. Moreover, as it has been
mentioned in Ref. [25]-[27], for normalized qubits the complex $1$-qubit, $2$%
-qubit and $3$-qubit are deeply related to division algebras via the Hopf
maps, $S^{3}\overset{S^{1}}{\longrightarrow }S^{2}$, $S^{7}\overset{S^{3}}{%
\longrightarrow }S^{4}$ and $S^{15}\overset{S^{7}}{\longrightarrow }S^{8}$,
respectively. So, according to this mathematical results (31) may only have
a reduced number of components:

\begin{equation}
\begin{array}{c}
ds^{8}=\xi _{1}g_{\mu _{1}}(x^{\sigma })dx^{\mu _{1}}+\xi _{2}g_{\mu _{1}\mu
_{2}}(x^{\sigma })dx^{\mu _{1}}dx^{\mu _{2}} \\ 
\\ 
+\xi _{4}g_{\mu _{1}...\mu _{4}}(x^{\sigma })dx^{\mu _{1}}...dx^{\mu
_{4}}+\xi _{8}g_{\mu _{1}...\mu _{8}}(x^{\sigma })dx^{\mu _{1}}...dx^{\mu
_{8}}.%
\end{array}
\tag{32}
\end{equation}

How may look the generalized Riemann curvature tensors for $g_{\mu
_{1}...\mu _{4}}(x^{\sigma })$? In Refs. [13]-[18] have already consider a
solution for this question. The general idea is to associate a four-rank
Riemann tensor for $g_{\mu _{1}...\mu _{4}}$. An alternative is to think
that one needs two derivatives for $g_{\mu _{1}...\mu _{4}}$. The first
derivative may lead to the analogue of the Christoffel connection and the
second to six-rank Riemann tensor. However, an interesting possibility
arises thinking about a kind of generalized vielbien $e_{\mu _{1}}^{a_{1}}$.
Recall that for a two-rank metric one may write

\begin{equation}
g_{\mu _{1}\mu _{2}}=e_{\mu _{1}}^{a_{1}}e_{\mu _{2}}^{a_{2}}\eta
_{a_{1}a_{2}}.  \tag{33}
\end{equation}%
So, the analogue of this expression for $g_{\mu _{1}\mu _{2}\mu _{3}\mu
_{4}} $ will be%
\begin{equation}
g_{\mu _{1}\mu _{2}\mu _{3}\mu _{4}}=e_{\mu _{1}\mu _{2}}^{a_{1}a_{2}}e_{\mu
_{3}\mu _{4}}^{a_{3}a_{4}}\eta _{a_{1}a_{2}a_{3}a_{4}}.  \tag{34}
\end{equation}%
Now, one may obtain a connection $\omega _{\mu _{1}\mu _{2}\mu
_{3}}^{a_{1}a_{2}a_{3}a_{4}}$ by properly combining a first derivative of $%
e_{\mu _{1}\mu _{2}}^{a_{1}a_{2}}$. In turn, this connection may lead to a
kind of Riemann tensor $\mathcal{R}\sim \partial \omega $ of the form $%
\mathcal{R}_{\mu _{1}\mu _{2}\mu _{3}\mu _{4}}^{a_{1}a_{2}a_{3}a_{4}}$.
Following with this development one may think on the dual of $\mathcal{R}%
_{\mu _{1}\mu _{2}\mu _{3}\mu _{4}}^{a_{1}a_{2}a_{3}a_{4}}$ as follows:

\begin{equation}
^{\ast }\mathcal{R}_{\mu _{1}\mu _{2}\mu _{3}\mu
_{4}}^{a_{1}a_{2}a_{3}a_{4}}=\frac{1}{4!}\varepsilon
_{a_{5}a_{6}a_{7}a_{8}}^{a_{1}a_{2}a_{3}a_{4}}\mathcal{R}_{\mu _{1}\mu
_{2}\mu _{3}\mu _{4}}^{a_{5}a_{6}a_{7}a_{8}}.  \tag{35}
\end{equation}%
Note that in this case an eight dimensional spacetime is necessary. In turn,
this expression may allow to consider the self (antiself) dual curvature

\begin{equation}
^{\pm }\mathcal{R}_{\mu _{1}\mu _{2}\mu _{3}\mu _{4}}^{a_{1}a_{2}a_{3}a_{4}}=%
\frac{1}{2}(\mathcal{R}_{\mu _{1}\mu _{2}\mu _{3}\mu
_{4}}^{a_{1}a_{2}a_{3}a_{4}}+\frac{1}{4!}\varepsilon
_{a_{5}a_{6}a_{7}a_{8}}^{a_{1}a_{2}a_{3}a_{4}}\mathcal{R}_{\mu _{1}\mu
_{2}\mu _{3}\mu _{4}}^{a_{5}a_{6}a_{7}a_{8}}).  \tag{36}
\end{equation}%
Thus, an action, analogue to the Ashtekar formalism, for $^{\pm }\mathcal{R}%
_{\mu _{1}\mu _{2}\mu _{3}\mu _{4}}^{a_{1}a_{2}a_{3}a_{4}}$ may be proposed,
namely

\begin{equation}
S=\int_{M^{8}}\varepsilon
_{a_{5}a_{6}a_{7}a_{8}}^{a_{1}a_{2}a_{3}a_{4}}e_{a_{1}a_{2}}^{\mu _{1}\mu
_{2}}e_{a_{3}a_{4}}^{\mu _{3}\mu _{4}\pm }\mathcal{R}_{\mu _{1}\mu _{2}\mu
_{3}\mu _{4}}^{a_{5}a_{6}a_{7}a_{8}},  \tag{37}
\end{equation}%
where $M^{8}$ is an appropriate eight dimensional manifold. For further
research one may be interested in developing the canonical quantization of
this action.

From the point of view of string theory one may even propose the action%
\begin{equation}
\begin{array}{c}
S=\int \sqrt{g}[g^{a_{1}a_{2}}\frac{\partial x^{\mu _{1}}}{\partial \sigma
^{a}}\frac{\partial x^{\mu _{2}}}{\partial \sigma ^{a}}\eta _{\mu _{1}\mu
_{2}}(x^{\sigma })+g^{a_{1}a_{2}}g^{a_{3}a_{4}}\frac{\partial x^{\mu _{1}}}{%
\partial \sigma ^{a_{1}}}\frac{\partial x^{\mu _{2}}}{\partial \sigma
^{a_{2}}}\frac{\partial x^{\mu _{3}}}{\partial \sigma ^{a_{3}}}\frac{%
\partial x^{\mu _{4}}}{\partial \sigma ^{a_{4}}}\eta _{\mu _{1}\mu _{2}\mu
_{3}\mu _{4}} \\ 
\\ 
+g^{a_{1}a_{2}}...g^{a_{n-1}a_{n}}\frac{\partial x^{\mu _{1}}}{\partial
\sigma ^{a_{1}}}\frac{\partial x^{\mu _{2}}}{\partial \sigma ^{a_{2}}}...%
\frac{\partial x^{\mu _{n}}}{\partial \sigma ^{a_{n}}}\eta _{\mu _{1}\mu
_{2}...\mu _{n}}].%
\end{array}
\tag{38}
\end{equation}%
Instead of $g^{a_{1}a_{2}}$ in (38), one can, of course, introduce a
generalized metric $g^{a_{1}a_{2}...}{}^{a_{n-1}a_{n}}$. But in this case,
instead of strings, the structure would correspond to a $n$-brane systems
(see Ref. [29] and references therein).

A rough explanation why we need a four-rank metric $g_{\mu _{1}...\mu
_{4}}(x^{\sigma })$ can be given by thinking from the perspective that in
the torus associated with (23) when $x=0$ one has two ellipses and therefore
eventually one will need two metrics which is precisely what it is happening
in the particular case when $g_{\mu _{1}\mu _{2}\mu _{3}\mu _{4}}$ is
splitted as $g_{\mu _{1}\mu _{2}\mu _{3}\mu _{4}}=g_{\mu _{1}\mu _{2}}g_{\mu
_{3}\mu _{4}}$.

Topologically, one may say that the Newtonian gravitational theory is
associated with a cone, while our proposed Spiric gravitational theory is
associated with a torus. Mathematically, however, the torus is part of a set
of Riemann surfaces. So, our approach may eventually be generalized to
arbitrary Riemann surfaces. In this case the topologically notions of Euler
characteristic and Pontryagin signature may be important mathematical
notions (see Ref. [28]).

In Ref. [30] it was discover that the Bernoulli lemniscate curve has a
hidden octahedron symmetry. This is achieved by first unifying the circle,
the hyperbola and the lemniscate by different maps and then using
stereographic projection to establish a link with the octahedron; one of the
Platonic solids. (It is worth mentioning that the octahedron is dual to the
cube.) Topologically the Platonic solids are homeomorphic to the sphere. So,
one may expect that these developments may eventually be connected with the
torus rather with the sphere. It will be a subject of another work to
investigate the connection between such a hidden octahedron symmetries and
the present formalism.

Moreover, the Bernoulli lemniscate arises naturally in superstring
quantization. So, one may expect that the present developments may be useful
to understand various aspects of such a quantization.

It worth mentioning that four-rank gauge field $c_{\mu _{1}\mu _{2}\mu
_{3}\mu _{4}}$ with the same symmetries as the Riemann tensor has been
proposed [31]. It seems interesting to explore a possible link between $%
c_{\mu _{1}\mu _{2}\mu _{3}\mu _{4}}$ and the four rank metric $g_{\mu
_{1}\mu _{2}\mu _{3}\mu _{4}}$. Since $c_{\mu _{1}\mu _{2}\mu _{3}\mu _{4}}$
arises from duality considerations of linearized gravity [32] one may expect
that such link may be obtained when one considers the linearized theory for $%
g_{\mu _{1}\mu _{2}\mu _{3}\mu _{4}}$.

Finally, from the point of view of graph theory the Bernoulli lemniscate may
be associated with the mathematical structure of oriented matroid theory
[33] which is a combinatorial structure that has been proposed as the
underlying mathematical framework for $M$-theory [34]. Since there are a
number of evidences that suggests that in fact $M$-theory is linked to
oriented matroid theory including connections with $p$-branes, qubit theory,
Chern-Simons theory, supergravity and string theory, among others (see Refs.
[34]-[38] and references therein) one wonders whether the present work may
be linked with these developments.

\bigskip \ 

\begin{center}
\textbf{Acknowledgments}
\end{center}

This work was partially supported by PROFAPI-UAS/2013.

\end{document}